\documentclass[12pt]{article}

\usepackage{amsmath}
\usepackage{amssymb}
\usepackage{latexsym}
\usepackage{graphicx}
\usepackage{graphics}

\newcommand{\sech}{\, {\mathrm {sech}}}

\begin{document}

\nopagebreak

\title{\textbf{Classical and thermodynamical aspects of black holes with conformally
coupled scalar field hair}\thanks{Talk given at the ``Workshop on Dynamics and Thermodynamics of Black Holes
and Naked Singularities'', Milan, May 2004.}}

\author{ {\Large {Elizabeth Winstanley}}\thanks{E-mail: E. Winstanley@sheffield.ac.uk} \\[0.2in]
{\it {Department of Applied Mathematics, The University of Sheffield,}} \\
{\it {Hicks Building, Hounsfield Road, Sheffield, S3 7RH, U.K.}} }

\maketitle

\bigskip
------------------------------------------------------------------------------------------------------
\begin{abstract}
\noindent
We discuss the existence, stability and classical thermodynamics of
four-dimensional, spherically symmetric black hole solutions of
the Einstein equations with a conformally coupled scalar field.
We review the solutions existing in the literature
with zero, positive and negative cosmological constant.
We also outline new results on the thermodynamics of these
black holes when the cosmological constant is non-zero.
\end{abstract}
\bigskip
\par------------------------------------------------------------------------------------------------------

\section{Introduction}
\label{sec:intro}

Black holes with a conformally coupled scalar field were first studied thirty years ago,
when Bekenstein \cite{bek74,bek75} rediscovered an exact solution of the Einstein equations
previously found by Bocharova et al \cite{bbm} but unknown in the West.
Although the much simpler case of a minimally coupled scalar field has been widely studied
in the literature over the intervening years (see, for example, \cite{heuslerbook} for a
review), interest in the conformally coupled case has been revived only comparatively
recently (for example, with the theorems of Saa \cite{saa1,saa2}).
Moreover, developments (particularly in string theory and higher-dimensional
``brane world'' models)
over the past few years have reignited relativists' interest in models with a
cosmological constant (both positive and negative), and in the last couple of years
new black hole solutions of the Einstein equations for
the conformally coupled scalar field system with a non-zero cosmological constant
have been found \cite{martinez,ewconf}.

As well as reviewing these developments, our other purpose in this note is
to begin a study of the thermodynamics of these black holes.
The presence of the non-minimal coupling of the scalar field to the space-time
curvature complicates the thermodynamics, particularly the entropy \cite{ashtekar}.

The outline of this paper is as follows.
In section \ref{sec:model} we introduce our model of a four-dimensional black hole
with a conformally coupled scalar field.
We also discuss a transformation due to Maeda \cite{maeda}, which converts this
system to a much simpler model of gravity with a minimally coupled scalar field.
Next, in section \ref{sec:bhsols}, we discuss the  solutions of the field equations
representing spherically symmetric black holes, considering the cases of zero, positive
and negative cosmological constant separately.
This involves reviewing the known solutions of Bekenstein \cite{bek74,bek75},
Bocharova et al \cite{bbm} and Martinez et al \cite{martinez} as well as
recent solutions in anti-de Sitter space found by the author \cite{ewconf}.
In addition to the linear stability properties of these solutions,
we also comment on uniqueness/non-existence theorems existing in the literature.
The thermodynamics of these solutions is the subject of section \ref{sec:thermo},
reviewing work by Zaslavskii \cite{zaslavskii} on the zero cosmological constant
case, and also outlining new work \cite{barlow} when the cosmological constant is non-zero.
Our conclusions are presented in section \ref{sec:conc}.

\section{The Model}
\label{sec:model}

We start with the following action, which describes a self-interacting scalar field
$\phi $ conformally coupled to gravity:
\begin{equation}
S=\int d^{4}x \, {\sqrt {-g}} \left[
\frac {1}{2}\left( R -2\Lambda \right)
-\frac {1}{2} \left( \nabla \phi \right) ^{2}
-\frac {1}{12}  R \phi ^{2} -V(\phi ) \right] .
\label{eq:action}
\end{equation}
Here $R$ is the Ricci scalar, $\Lambda $ the cosmological constant,
$V(\phi )$ the scalar field self-interaction potential and
$\left( \nabla \phi \right) ^{2} = \nabla _{\mu } \phi \nabla ^{\mu } \phi $.
The metric has signature $(-+++)$ and we use units in which $\hbar =c=8\pi G=k_{B}=1$.
Variation of the action (\ref{eq:action}) gives the field equations
\begin{eqnarray}
\left[ 1- \frac {1}{6} \phi ^{2} \right] G_{\mu \nu }
+g_{\mu \nu } \Lambda
& = &
\frac {2}{3} \nabla _{\mu } \phi \nabla _{\nu }
\phi
-  \frac {1}{6} g_{\mu \nu } \left(
\nabla \phi \right)^{2}
-\frac {1}{3} \phi \nabla _{\mu } \nabla _{\nu } \phi
+\frac {1}{3} g_{\mu \nu } \phi \nabla ^{\rho }\nabla _{\rho } \phi
\nonumber \\
& &
- g_{\mu \nu } V(\phi ) ;
\label{eq:Einstein}
\\
\nabla _{\mu } \nabla ^{\mu } \phi & = & \frac {1}{6}  R\phi +
\frac {dV}{d \phi } .
\label{eq:phi}
\end{eqnarray}
Taking the trace of the Einstein equation (\ref{eq:Einstein}) and using the
scalar field equation (\ref{eq:phi}) to substitute for
$\nabla ^{\mu }\nabla _{\mu } \phi $, the Ricci scalar takes the form:
\begin{equation}
R= 4\Lambda + 4V(\phi ) - \phi \frac {dV}{d\phi }.
\label{eq:Ricci}
\end{equation}
Here we are concerned with four-dimensional, spherically symmetric
black holes in which the metric takes the form
\begin{displaymath}
ds^{2}  =  -N(r) \exp (2\delta (r)) \, dt^{2}
+ N(r) ^{-1} dr^{2}
+ r^{2} \, d\theta ^{2}
+ r^{2} \sin ^{2} \theta \, d \varphi ^{2},
\end{displaymath}
where
\begin{displaymath}
N(r)= 1 - \frac {2m(r)}{r} - \frac {\Lambda r^{2}}{3},
\end{displaymath}
and we assume that the scalar field $\phi $ depends only on the
radial co-ordinate $r$.
The field equations then take the form (a prime ${}'$ denotes $d/dr$):
\begin{eqnarray}
\frac {2}{r^{2}} \left( 1- \frac {1}{6} \phi ^{2} \right) m'  & = &
\frac  {1}{6} \Lambda \phi ^{2} +\frac {1}{6} N \phi '^{2}
-\frac {1}{6} \phi \phi ' N' - \frac {1}{3} N \phi \phi ''
-\frac {2N}{3r}  \phi \phi '
+ V(\phi ) ;
\nonumber \\
\frac {2}{r} \left( 1 -\frac {1}{6} \phi ^{2} \right) \delta ' & = &
\frac {2}{3} \phi '^{2} - \frac {1}{3} \phi \phi ''
+ \frac {1}{3} \phi \phi ' \delta '  ;
\nonumber \\
0 & = &
N \phi '' + \left( N \delta ' + N' + \frac {2N}{r} \right)
\phi ' - \frac {1}{6} R \phi - \frac {dV}{d \phi } .
\label{eq:phisym}
\end{eqnarray}
The conformal coupling of the scalar field to the geometry means that the field
equations (\ref{eq:phisym}) are considerably more complex than the corresponding
equations for a minimally coupled scalar field.
Numerical integration of (\ref{eq:phisym}) can be simplified by first eliminating
the Ricci scalar from the scalar field equation using (\ref{eq:Ricci}) and
then using the scalar field equation to eliminate $\phi ''$ from the Einstein
equations.
Here we are interested in black hole solutions of the field equations, with
a regular, non-extremal, event horizon at some value of the radial co-ordinate,
$r=r_{h}$.
At $r=r_{h}$, the metric function $N(r)$ will have a simple zero, and we
assume that all the field variables have Taylor expansions about this point.
If the cosmological constant is positive, there will also be a regular cosmological
horizon at a distinct value of $r=r_{c}$, where again $N$ has a simple zero.
If $\Lambda $ is zero or negative, there will be no zeros of $N$ for $r$ larger
than $r_{h}$.
We also assume that the scalar field tends to a constant limit as $r\rightarrow \infty $,
sufficiently quickly that the metric approaches flat Minkowski, de Sitter or anti-de Sitter
space as $\Lambda $ is zero, positive or negative respectively.
The boundary conditions in anti-de Sitter space are discussed in more detail in
\cite{ewconf}; however, the consequences of the behaviour at infinity for defining
the mass and other conserved quantities are rather complex, and we will shall not
go into this further here.

If we assume that $\phi ^{2} \neq 1/6$ everywhere, then the system described
by the action (\ref{eq:action}) can be subject to a conformal transformation
\cite{maeda}:
\begin{equation}
{\bar {g}}_{\mu \nu } =\Omega g_{\mu \nu } ,
\label{eq:metricT}
\end{equation}
where
\begin{equation}
\Omega = 1- \frac {1}{6} \phi ^{2} .
\label{eq:Omega}
\end{equation}
Under this transformation, the action (\ref{eq:action}) becomes that of a
minimally coupled scalar field:
\begin{equation}
S=\int d^{4}x \, {\sqrt {-{\bar {g}}}} \left[
\frac{1}{2} \left( {\bar {R}}-2\Lambda  \right)
-\frac {1}{2} \left( {\bar {\nabla }} \Phi \right) ^{2}
-U(\Phi ) \right] ,
\label{eq:actionT}
\end{equation}
where a bar denotes quantities calculated using the transformed metric
(\ref{eq:metricT}); the new scalar field $\Phi $ is defined as \cite{maeda}:
\begin{equation}
\Phi = {\sqrt {6}} \tanh ^{-1} \left( \frac {\phi }{{\sqrt {6}}}
\right) ;
\label{eq:Phi}
\end{equation}
and the transformed potential $U(\Phi )$ takes the form \cite{ewconf}:
\begin{equation}
U(\Phi ) = \frac {V(\phi )+\frac {1}{6} \Lambda  \phi ^{2} \left(
2- \frac {1}{6} \phi ^{2} \right) }{ \left( 1- \frac {1}{6} \phi ^{2} \right) ^{2}} .
\label{eq:potT}
\end{equation}
The transformed metric (\ref{eq:metricT}) will also be spherically symmetric, and
we define a new radial co-ordinate ${\bar {r}}$ by:
\begin{equation}
{\bar {r}} = \left( 1 - \frac {1}{6} \phi ^{2} \right) ^{\frac {1}{2}} r,
\label{eq:rbar}
\end{equation}
so that (\ref{eq:metricT}) takes the form
\begin{displaymath}
d{\bar {s}}^{2} = -{\bar {N}}({\bar {r}})
e^{2{\bar {\delta }}({\bar {r}})} dt ^{2}
+ {\bar {N}}({\bar {r}}) ^{-1} d{\bar {r}}^{2}
+ {\bar {r}}^{2} \left( d\theta ^{2} +
\sin \theta ^{2} \, d\varphi ^{2} \right) ,
\end{displaymath}
where
\begin{eqnarray}
{\bar {N}} & = &
N \left( 1- \frac {1}{6} \phi ^{2} -\frac {1}{6} r \phi \phi' \right) ^{2}
\left( 1- \frac {1}{6} \phi ^{2} \right) ^{-2}  ;
\nonumber \\
e^{\bar {\delta }} & = & e^{\delta }
\left( 1- \frac{1}{6} \phi ^{2} -\frac {1}{6} r \phi \phi' \right) ^{-1}
\left( 1- \frac {1}{6} \phi ^{2} \right) ^{\frac {3}{2}} ;
\label{eq:metricbar}
\end{eqnarray}
and
\begin{displaymath}
{\bar {N}}({\bar {r}}) = 1
- \frac {2{\bar {m}}({\bar {r}})}{{\bar {r}}}
-\frac {\Lambda {\bar {r}}^{2}}{3}.
\end{displaymath}
The great advantage of using the conformal transformation (\ref{eq:metricT})
is that the field equations derived from the new action (\ref{eq:actionT}) take
the much simpler form:
\begin{eqnarray}
\frac {d{\bar {m}}}{d{\bar {r}}} & = &
\frac {{\bar {r}}^{2}}{4}  {\bar {N}}
\left(\frac {d\Phi }{d{\bar {r}}}\right) ^{2}
+ \frac {{\bar {r}}^{2}}{2}  U(\Phi ) ;
\nonumber \\
\frac {d{\bar {\delta }}}{d{\bar {r}}} & = &
\frac {{\bar {r}}}{2} \left( \frac {d\Phi }{d{\bar {r}}} \right) ^{2} ;
\nonumber \\
0 & = &
{\bar {N}} \frac {d^{2}\Phi }{d{\bar {r}}^{2}}
+ \left( {\bar {N}} \frac {d{\bar {\delta }}}{d{\bar {r}}}
+ \frac {d{\bar {N}}}{d{\bar {r}}}
+ \frac {2{\bar {N}}}{{\bar {r}}} \right)
\frac {d\Phi }{d{\bar {r}}}
- \frac {dU}{d\Phi } .
\label{eq:mincscal}
\end{eqnarray}
The relations (\ref{eq:metricbar}) ensure that the horizon structure of the geometry
is maintained after the conformal transformation, and, for the solutions we find using
this transformation in section \ref{sec:Lneg}, this is true also for the behaviour at
infinity.
One has to be careful in making use of the conformal transformation to ensure that
in fact it is valid everywhere (i.e. the conformal factor $\Omega $ (\ref{eq:Omega})
does not vanish).
In reference \cite{ashtekar}, it is argued that the vanishing of $\Omega $ at
a single point is not catastrophic for having a well-defined theory; however
it is imperative that $\Omega $ must not vanish on an open set.

\section{Classical Black Hole Solutions}
\label{sec:bhsols}

In this section we discuss the existence and stability of black hole
solutions of the field equations (\ref{eq:Einstein}-\ref{eq:phi}),
considering zero, positive and negative cosmological constant in turn.

\subsection{$\Lambda =0$}
\label{sec:Lzero}

For zero cosmological constant, there is a famous solution of the field equations
(\ref{eq:Einstein}-\ref{eq:phi}) due to Bocharova et al \cite{bbm} and
discovered independently by Bekenstein \cite{bek74,bek75}, known as the
BBMB black hole \cite{bek96}.
In this case the scalar field self-interaction potential vanishes, and the metric and
scalar field functions take the form:
\begin{equation}
N(r) = \left( 1 - \frac {M}{r} \right) ^{2}, \qquad \qquad
\delta (r) \equiv 0, \qquad \qquad
\phi = \frac {1}{{\sqrt {6}}} \frac {M}{r-M}.
\label{eq:BBMB}
\end{equation}
The solution (\ref{eq:BBMB}) represents a neutral black hole, although it is possible
to couple the geometry also to an electromagnetic field \cite{bek74,bek75}.
The metric (\ref{eq:BBMB}) is that of an extremal Reissner-Nordstr\"om black hole,
even when there is no electromagnetic field.
The inclusion of the scalar field does not introduce any additional parameters
into the solution of the field equation, so there are no extra degrees of freedom
and the ``hair'' of the scalar field is sparse.
The BBMB solution has also been controversial because the scalar field diverges at the
event horizon, where $r=M$.
Sudarsky and Zannias \cite{zannias} have argued that this makes the solution unphysical,
although Bekenstein \cite{bek75} maintains that a particle coupled to the scalar field
would experience nothing pathological at the event horizon, where, it must be stressed,
the geometry is perfectly regular.
Both the neutral and charged BBMB black holes have been shown to be unstable \cite{bronnikov}.
The thermodynamical properties of the BBMB black holes will be discussed in
section~\ref{sec:BBMB}.

In the 1990's Xanthopoulos and Zannias \cite{xanth} proved that the BBMB black hole
is the unique static, asymptotically flat, solution of the Einstein equations with
a conformally coupled scalar field and zero potential.
For more general potentials, no non-trivial solutions are known, and it is unlikely
that any exist.
However, there is at present no complete proof of this statement.
A very general paper by Bekenstein and Mayo \cite{mayo} proves that there
can be no charged solutions; however, their proof does not extend to neutral black holes
with a conformally coupled scalar field (although they do rule out solutions for some
other non-minimal scalar field couplings).
There are related results by Saa \cite{saa1,saa2}, which rely on fairly strong conditions
on the scalar field; these essentially mean that the conformal transformation
(\ref{eq:metricT}) is valid everywhere so that the conformally coupled system can be
studied via the minimally coupled system (\ref{eq:mincscal}), which is much easier to
work with, and for which there are general no-hair theorems (see, for example,
\cite{bek72}--\cite{sudarsky}).
Of course, this leaves open the question of whether there are solutions for which
$\phi ^{2}=6$ for some value of $r$, although there is comprehensive numerical evidence
that this cannot happen, at least for zero potential \cite{pena}.

\subsection{$\Lambda >0$}
\label{sec:Lpos}

For a positive cosmological constant, a very simple proof
(following \cite{bek72}) suffices to show that
there are no non-trivial black hole solutions if the potential is either zero or
represents a massive scalar field with no additional self-interaction:
\begin{equation}
V (\phi ) = \frac {1}{2} \mu ^{2} \phi ^{2},
\label{eq:potential}
\end{equation}
where $\mu $ is the mass of the scalar field.
The proof can be applied directly to the conformally coupled system, without needing
to transform to the minimally coupled system \cite{ewconf}.
Take the scalar field equation (\ref{eq:phisym}), substitute for the Ricci scalar from
(\ref{eq:Ricci}), and the potential (\ref{eq:potential}), multiply both sides of the
equation by $\phi r^{2} e^{\delta }$ and then integrate from the black hole event
horizon $r=r_{h}$ to the cosmological horizon $r=r_{c}$:
\begin{equation}
0 =
\int _{r_{h}}^{r_{c}} dr
\, r^{2} e^{\delta } \left[
\left( \frac {2}{3} \Lambda + \mu ^{2}
+\frac {1}{6} \mu ^{2} \phi ^{2}  \right) \phi ^{2}
+ N  \phi  ^{'2} \right]
-\left[ N r^{2} e^{\delta }
 \phi \phi '
\right] _{r_{h}}^{r_{c}} ,
\label{eq:posdefbit}
\end{equation}
where we have integrated by parts to yield the boundary term.
If we assume that the event and cosmological horizons are regular and the scalar
field is also regular there, then the boundary term vanishes.
The integrand in (\ref{eq:posdefbit}) is then manifestly positive definite, and
the only possibility is that $\phi ,\phi '\equiv 0$, giving the Schwarzschild-de Sitter
black hole.
The assumption that the scalar field is regular on the horizons means that this method
does not apply to the BBMB solution.

The above proof does not readily extend to more general potentials.
For a quartic potential
\begin{displaymath}
V(\phi ) = \alpha \phi ^{4},
\end{displaymath}
where $\alpha $ is a coupling constant, there is a non-trivial solution
due to Martinez et al \cite{martinez}, which we shall refer to as the MTZ black hole.
This solution is the analogue of the BBMB black hole in the presence of a
positive cosmological constant.
The neutral black hole solution exists only for
$\alpha = -\Lambda / 36$
and has metric and scalar field functions
\begin{equation}
N(r)= \left( 1 - \frac {M}{r} \right) ^{2} -\frac {\Lambda r^{2}}{3}, \qquad \qquad
\delta (r) \equiv 0, \qquad \qquad
\phi (r) = \frac {1}{{\sqrt {6}}} \frac {M}{r-M}.
\label{eq:MTZ}
\end{equation}
Like the BBMB solution, there are also charged MTZ black holes, which exist for all
values of $\alpha $ less than $-\Lambda /36$.
The metric for these is the same as that above (\ref{eq:MTZ}), but the scalar
field in this case is given by
\begin{equation}
\phi (r) = {\sqrt {6}} \frac {{\sqrt {M^{2}-Q^{2}}}}{r-M} =
{\sqrt {-\frac {\Lambda }{6\alpha }}} \frac {M}{r-M},
\label{eq:MTZcharged}
\end{equation}
where the charge $Q$ of the black hole is related to the mass $M$ by
\begin{equation}
\left( \frac {Q}{M} \right) ^{2} = 1 + \frac {\Lambda }{36\alpha }.
\label{eq:QMrel}
\end{equation}
It should be emphasized that the charge $Q$ is not equal to the mass $M$ as might be
expected from looking at the metric (\ref{eq:MTZ}).
Indeed, the metric has the form of Reissner-Nordstr\"om-de Sitter space even when
there is no electromagnetic field and therefore no charge present.

The geometry of the MTZ black holes
has a non-extremal event horizon and a cosmological horizon.
However, if $M>{\sqrt {3/16\Lambda }}$ or $\Lambda <0$,
the metric (\ref{eq:MTZ}) represents a naked
singularity.
Unlike the BBMB black hole, the scalar field (\ref{eq:MTZ}) is regular on both horizons,
and has a pole which lies inside the event horizon.
This means that the solution cannot immediately be ruled out as ``unphysical'', but, like
the BBMB black hole, the scalar field does not really introduce additional ``hair'' as
there are no new parameters involved.
In section \ref{sec:MTZ} we shall discuss the thermodynamic properties of the MTZ black
holes.

\begin{figure}
\begin{center}
\includegraphics[height=4in,angle=270]{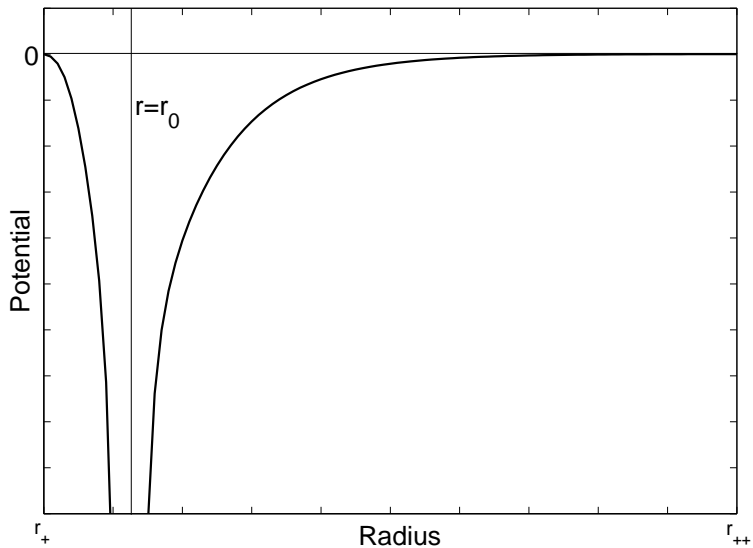}
\caption{Graph of the perturbation potential ${\cal {U}}$ for the neutral
MTZ black holes, and most of the charged MTZ black holes.
The potential has a negative double pole, from which it can
be deduced that the black holes are unstable.}
\label{fig:potential}
\end{center}
\end{figure}

The stability of these black holes under linear, spherically symmetric perturbations
was studied in \cite{harper}.
Standard techniques are used to write the field equations as a single equation
for a perturbation $\Psi $, which is related to the perturbation of the scalar field
$\delta \phi $ by:
\begin{displaymath}
\Psi = r \left|
1 - \frac {1}{6} \phi ^{2}
\right| ^{-\frac {1}{2}} \delta \phi ,
\end{displaymath}
giving the perturbation equation in the standard Schr\"odinger form
\begin{equation}
-\frac {\partial ^{2} \Psi }{\partial x_{*}^{2}} +
{\cal {U}} \Psi = \sigma ^{2} \Psi ;
\label{eq:pert}
\end{equation}
where $x_{*}$ is the usual ``tortoise'' co-ordinate defined by
\begin{displaymath}
\frac {dx_{*}}{dr} = \frac {1}{N};
\end{displaymath}
${\cal {U}}$ is the perturbation potential, whose form can be found explicitly in
\cite{harper}; and $\sigma ^{2}$ is the eigenvalue, such that the equilibrium
configurations are unstable if there are bound state solutions of (\ref{eq:pert})
with $\sigma^{2}<0$.
For all the neutral MTZ black holes, the perturbation potential ${\cal {U}}$ has the
form shown in figure \ref{fig:potential}, with a negative double pole at some value
of $r$ between the event and cosmological horizons.
Standard results in quantum mechanics can then be used to deduce that there are
unstable bound state solutions of (\ref{eq:pert}) with $\sigma ^{2}<0$ for this
type of potential, so that all the neutral MTZ black holes are unstable.

\begin{figure}
\begin{center}
\includegraphics[height=4in,angle=270]{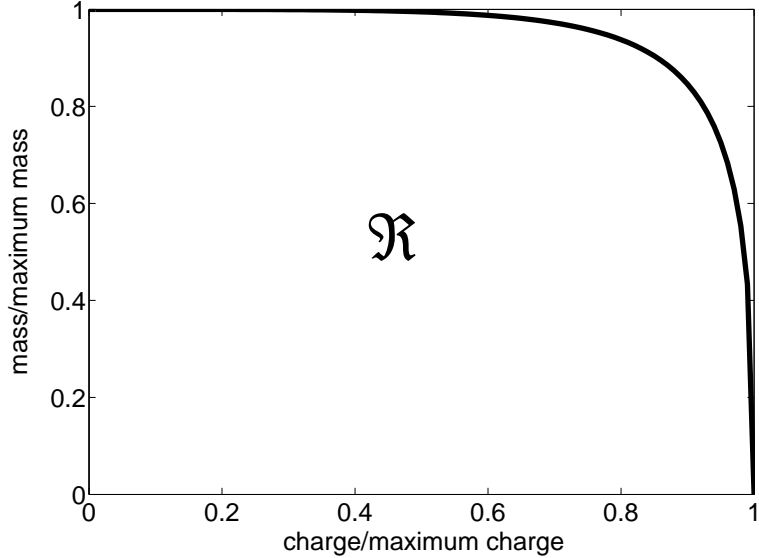}
\caption{The parameter space of the charged MTZ black holes. For
black holes corresponding to points inside the region $\Re $, the perturbation
potential has the form shown in figure {\protect {\ref{fig:potential}}}, and analytic techniques
suffice to show that the black holes are unstable.
For black holes corresponding to points outside the region $\Re $, numerical techniques
can be employed to show instability \cite{harper}.}
\label{fig:param}
\end{center}
\end{figure}

The situation for the charged black holes is more complex.
The parameter space for the charged black holes can be divided into two regions, as shown
in figure \ref{fig:param}.
For black holes corresponding to points lying inside the region $\Re $ in figure
\ref{fig:param}, which makes up most of the parameter space, the perturbation
potential ${\cal {U}}$ has the form shown in figure \ref{fig:potential},
and the same techniques as applied to the neutral MTZ black holes can be used
to show that these charged black holes are also unstable.
For black holes corresponding to points outside $\Re $, the potential does not possess
a pole and numerical techniques have to be used.
However, for all the configurations tested, the black holes were found to be unstable (see
\cite{harper} for details).

The question of the existence of solutions for other scalar field potentials remains open
when the cosmological constant is positive.
For minimally coupled scalar fields in the presence of a positive cosmological constant,
it is known that non-trivial solutions exist, for example, for the Higgs-like potential
\cite{toriidS}, and it may well be that analogues of these solutions exist for
conformally coupled scalar fields.
However, one may conjecture that such solutions, like the MTZ black holes and those
with a minimally coupled scalar field \cite{toriidS}, would be unstable.

\subsection{$\Lambda <0$}
\label{sec:Lneg}

The proof employed in the previous subsection to show that there are no non-trivial
solutions when $\Lambda >0$ for simple potentials (\ref{eq:potential}) can easily be
seen to fail when the cosmological constant is negative.
At least for $\mu ^{2}< -2\Lambda /3$, it is no longer the case that the integrand
in (\ref{eq:posdefbit}) is positive definite (furthermore, it can be shown in this
case that the boundary term in (\ref{eq:posdefbit}) also does not vanish).
This leaves open the possibility of non-trivial solutions in this case.
Such solutions were found numerically in \cite{ewconf} and discussed in detail in that work.
Here we simply present a brief summary of the main features.

\begin{figure}
\begin{center}
\includegraphics[height=4in]{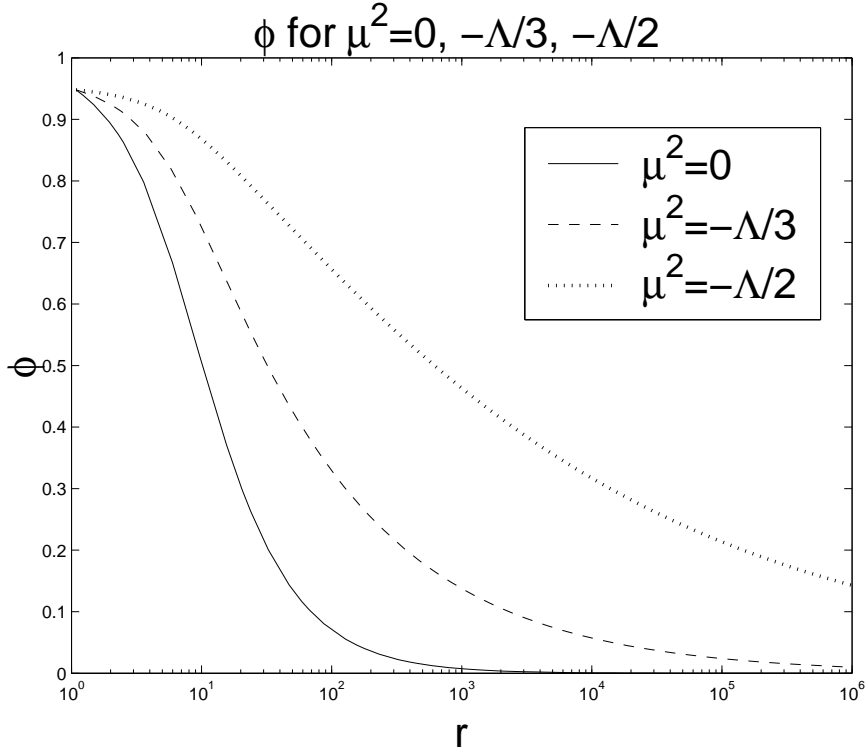}
\caption{Graph of $\phi $ (vertical axis) against $r$ (horizontal axis),
for $\Lambda = -0.1$, ${\bar {r}}_{h}=1$ and $\Phi ({\bar {r}}_{h})=1$, and
three values of the scalar field mass.
The event horizon is located at the left-hand edge of the graph.
The scalar field is monotonic and has no zeros.}
\label{fig:littlephi}
\end{center}
\end{figure}

Firstly, analysis of the conformally coupled scalar field equations (\ref{eq:phisym})
reveals that the scalar field is divergent at infinity if $\mu ^{2}>-2\Lambda /3$;
therefore we shall consider only values of $\mu ^{2}$ less than $-2\Lambda /3$ (including
the case $\mu ^{2}=0$).
The numerical solutions are generated using the conformal transformation (\ref{eq:metricT})
as the minimally coupled scalar field equations (\ref{eq:mincscal}) are easier to implement
than the conformally coupled scalar field equations (\ref{eq:phisym}).
With the form of the self-interaction potential (\ref{eq:potential}), the transformed potential
$U(\Phi )$ (\ref{eq:potT}) takes the form
\begin{displaymath}
U(\Phi ) = \Lambda
\left[  1+  \cosh ^{2} \left( \frac {\Phi }{{\sqrt {6}}} \right)
\right]
\sinh ^{2} \left( \frac {\Phi }{{\sqrt {6}}} \right)
+ 3 \mu ^{2} \sinh ^{2} \left( \frac {\Phi }{{\sqrt {6}}} \right)
\cosh ^{2} \left( \frac {\Phi }{{\sqrt {6}}} \right) .
\end{displaymath}
The field equations (\ref{eq:mincscal}) are then integrated with this potential from
the event horizon ${\bar {r}}={\bar {r}}_{h}$ out to large ${\bar {r}}$.
Solutions are found for a continuum of values of $\Phi $ at the event horizon,
with possibly an upper bound on the modulus of $\Phi $ at the event horizon.
The minimally coupled scalar field $\Phi $ and corresponding metric functions
${\bar {N}}$ and ${\bar {\delta }}$ are then transformed back to the conformally
coupled scalar field $\phi $ and original metric variables $N$ and $\delta $, using
(\ref{eq:Phi}) and the relations
\begin{eqnarray*}
{\bar {N}} & = &
N \left[ 1 + \frac {1}{{\sqrt {6}}} {\bar {r}}
\frac {d\Phi }{d{\bar {r}}} \tanh \left(
\frac {\Phi }{{\sqrt {6}}} \right) \right] ^{-2} ;
\nonumber \\
e^{{\bar {\delta }}} & = &
e^{\delta } \left[ 1 + \frac {1}{{\sqrt {6}}} {\bar {r}}
\frac {d\Phi }{d{\bar {r}}} \tanh \left(
\frac {\Phi }{{\sqrt {6}}} \right) \right]
\sech \left( \frac {\Phi }{{\sqrt {6}}} \right)  ;
\end{eqnarray*}
taking into account the change in the radial variable (\ref{eq:rbar}).
Typical solutions are shown in figure \ref{fig:littlephi}, where we
plot the conformally coupled scalar field $\phi$  for
the particular values $\Lambda = -0.1$, ${\bar {r}}_{h}=1$,
$\Phi ({\bar {r}}_{h})=1$, and three values of the mass: $\mu ^{2}=0$,
$\mu ^{2}= - \Lambda /3$ and $\mu ^{2}=-\Lambda /2$.
Similar results were found for other values of the parameters and initial conditions.
In each case we find that the scalar field monotonically decreases from its value
on the event horizon and decays to zero at infinity.
The rate of decay is slower for larger mass, and in each case the scalar field has no zeros.
The thermodynamic properties of these solutions are discussed in section \ref{sec:adS}.

\begin{figure}
\begin{center}
\includegraphics[height=4in]{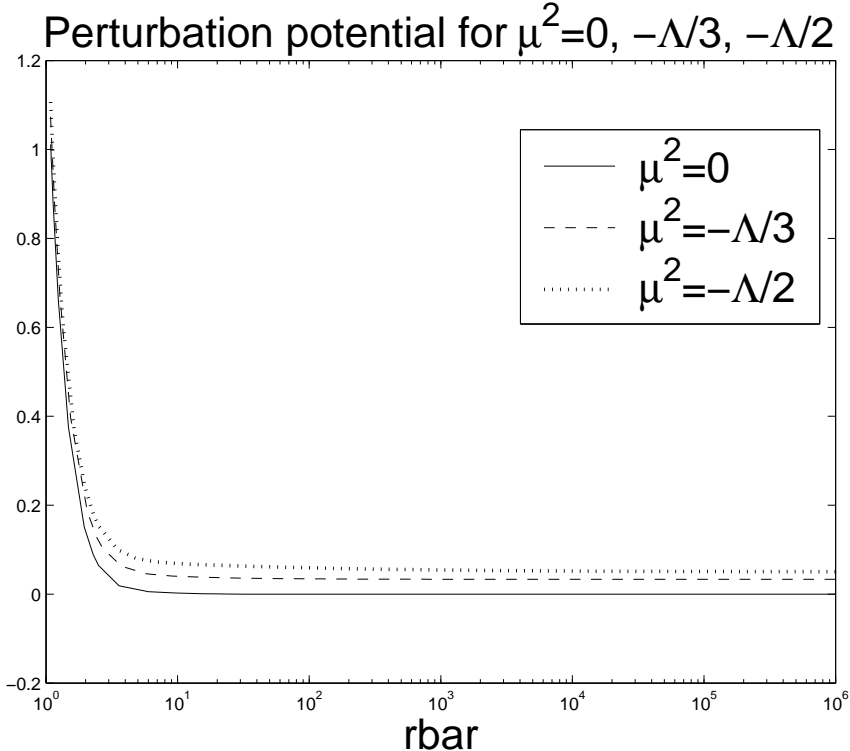}
\caption{Graph of the perturbation potential
${\cal {U}} {\bar {r}}^{2} e^{-2{\bar {\delta }}} {\bar {N}}^{-1}$
for the solutions plotted
in figure \ref{fig:littlephi}.
The independent variable is ${\bar {r}}$, defined by (\ref{eq:rbar}).
The potential is everywhere positive, from which it can be deduced that
these solutions are linearly stable.
Note that in each case the potential ${\cal {U}}$ vanishes at the
event horizon because we have divided by a factor of ${\bar {N}}$.}
\label{fig:AdSstab}
\end{center}
\end{figure}

Various properties of these solutions are discussed in \cite{ewconf}, including a
linear stability analysis under spherically symmetric perturbations.
The stability analysis proceeds similarly to that for the MTZ black holes in section
\ref{sec:Lpos}, except that for at least some of the solutions, the perturbation
potential ${\cal {U}}$
in the perturbation equation (\ref{eq:pert})
is everywhere positive (see figure \ref{fig:AdSstab}).
From the form of the potential in figure \ref{fig:AdSstab},
it can be deduced that these solutions are linearly stable.

It remains to be investigated whether similar solutions exist for different scalar
field potentials, but we do not anticipate such solutions to be fundamentally different
from those exhibited here.
A more interesting open question is whether there exist solutions which are not
conformally related to solutions of the minimally coupled scalar field equations,
presumably because $\phi ^{2}=6$ at some point so that the conformal transformation
(\ref{eq:metricT}) is not valid everywhere.

\section{Thermodynamics}
\label{sec:thermo}

Having reviewed the existence and stability of hairy black hole solutions
of the Einstein equations with a conformally coupled scalar field, we now turn to the
other topic in the talk, namely their thermodynamic properties.
The presence of the conformally coupled scalar field means that the thermodynamics of
these black holes differs from standard black hole thermodynamics, even when, as in the
case of the BBMB (\ref{eq:BBMB}) and MTZ (\ref{eq:MTZ}) black holes, the geometry
is actually that of a conventional black hole.

The temperature of the black holes is unaffected by the conformally coupled scalar
field, however, and given by the usual Hawking formula:
\begin{equation}
T = \frac {1}{4\pi } N' (r_{h}) e^{\delta (r_{h})}.
\label{eq:temp}
\end{equation}
On the other hand, the entropy of the black hole is no longer given by the usual area
formula, but acquires a multiplicative factor due to the conformally coupled scalar
field:
\begin{equation}
S = \pi r_{h}^{2} \left( 1- \frac {1}{6} \phi (r_{h})^{2} \right) .
\label{eq:entropy}
\end{equation}
There are a number of ways of deriving this result.
Firstly, there is the approach due to Iyer and Wald \cite{iyer}, where the entropy of a black
hole in a theory with a very general action is derived using a Noether current approach to
the first law.
This method gives the entropy to be:
\begin{displaymath}
S= -2 \pi  \int _{\Sigma } E^{\mu \nu \alpha \beta } \epsilon _{\mu \nu }
\epsilon _{\alpha \beta } ;
\end{displaymath}
where the integration is performed over the bifurcation two-sphere of the event horizon
$\Sigma $, with binormal $\epsilon _{\mu \nu }$, and $E^{\mu \nu \alpha \beta }$ is
the functional derivative of the Lagrangian $L$ of the theory with respect to
$R_{\mu \nu \alpha \beta }$, treated as a variable independent of the metric $g_{\mu \nu }$.
Applying this formula to our action (\ref{eq:action}) gives the entropy (\ref{eq:entropy}).
The entropy (\ref{eq:entropy}) has also been rederived recently using the isolated horizons approach
to deriving the first law \cite{ashtekar}.
Furthermore, if the conformal transformation (\ref{eq:metricT}) is valid, it is clear
that the entropy of the black hole in the minimally coupled scalar field system should simply
be one quarter the area of the event horizon.
However, because we need to change the radial co-ordinate in this transformation
(\ref{eq:rbar}), this entropy is in fact equal to (\ref{eq:entropy}) in terms of quantities
in the original, conformally coupled system.
Assuming that the corresponding black holes in the two systems have the same entropy (since
they are different descriptions of the same thing) gives the formula (\ref{eq:entropy})
for the entropy of the black hole with a conformally coupled scalar field.

In this section we will calculate the temperature and entropy for the various black hole
solutions discussed in section \ref{sec:bhsols}, taking the cases $\Lambda $ zero, positive
and negative in turn.

\subsection{Thermodynamics of the BBMB Black Hole}
\label{sec:BBMB}

Since it is an extremal black hole, the BBMB solution (\ref{eq:BBMB}) has zero
Hawking temperature (\ref{eq:temp}).
However, the scalar field diverges at the event horizon and so its entropy (\ref{eq:entropy})
is formally infinite.
The thermodynamics of this black hole, along with others with zero temperature
and infinite entropy,
has been studied in some
detail by Zaslavskii \cite{zaslavskii}.
It was found that for black holes with infinite horizon area (arising, for example,
in Brans-Dicke theory) it is possible to place an additional inner boundary on
the event horizon of the black hole in the usual Euclidean action approach and
thereby derive a finite, well-defined value of the black hole entropy.
However, this technique failed to give a sensible answer for the BBMB black hole
and Zaslavskii concluded that the BBMB black hole is not an object with a
conventional thermodynamic interpretation.
It remains to be seen whether there is an alternative approach to black hole thermodynamics
(such as that of \cite{fatibene}) which leads to a consistent interpretation of the
BBMB black hole.

\subsection{Thermodynamics of the MTZ Black Hole}
\label{sec:MTZ}

The thermodynamics of the MTZ black holes discussed in section \ref{sec:Lpos}
is particularly interesting.
Here we will calculate temperature and entropy for both the neutral and charged MTZ
black holes and comment on their properties.
More detailed thermodynamic properties of the MTZ black holes will be
described elsewhere \cite{barlow}.

The MTZ black holes have a metric which is identical to that of a special case of
Reissner-Nordstr\"om-de Sitter, which has been studied by Mellor and Moss \cite{mellor1,mellor2}.
The metric has two horizons, an event horizon and a cosmological horizon, but they
have the same temperature in this special case, given by
\begin{equation}
T= \frac {1}{2\pi l} {\sqrt {1-\frac {4M}{l}}} ,
\label{eq:MTZtemp}
\end{equation}
where $l^{2}=3/\Lambda $.
The entropy however, is rather more complicated and we need to consider the neutral and
charged MTZ black holes separately.

\begin{figure}
\begin{center}
\includegraphics[width=10cm,angle=270]{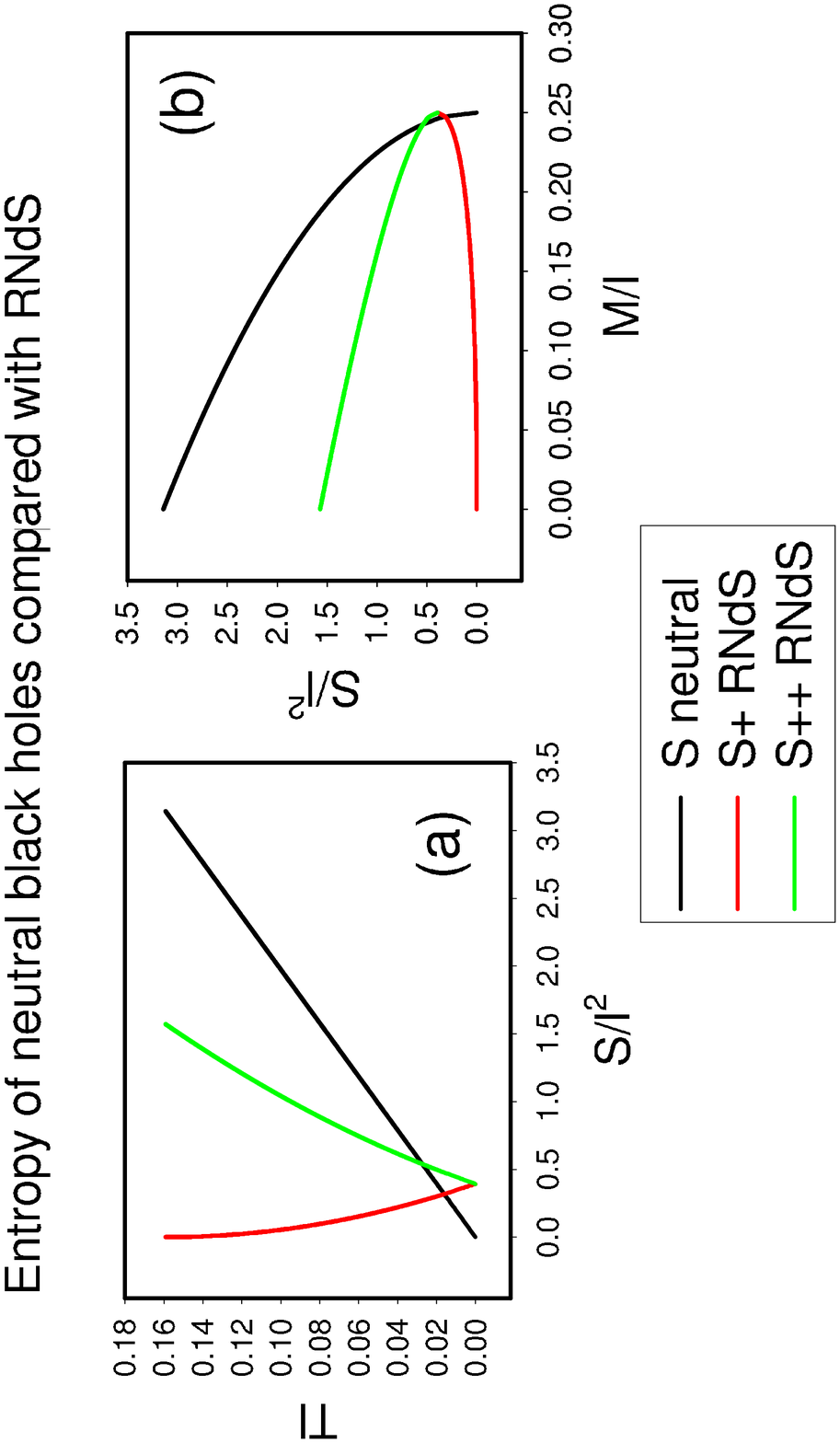}
\caption{Graphs of  (a) temperature against cosmological horizon entropy
and (b) cosmological horizon entropy against mass for the
neutral MTZ black holes,
compared with the entropies of the event ($S{+}$) and cosmological
($S{++}$) horizons for
Reissner-Nordstr\"om-de Sitter (RNdS) with the same metric but no scalar field.}
\label{fig:mtzneut}
\end{center}
\end{figure}

Firstly, we consider the neutral black holes.
In this case the entropy of the event horizon (which we denote by
$S_{h}$)  is negative,
whereas that of the cosmological
horizon ($S_{c}$) is positive, given by
\begin{displaymath}
S_{c}=-S_{h}= \pi l^{2} {\sqrt {1 - \frac {4M}{l}}}
=2 \pi ^{2} l^{3} T.
\end{displaymath}
The total entropy of these black holes, constructed by simply adding the entropies
of the event and cosmological horizons, is therefore zero, even though the temperature
is non-zero and finite.
This may indicate some form of thermodynamic instability, in addition to the classical
instability discussed in section \ref{sec:Lpos}.
The conformally coupled scalar field therefore has a considerable effect on the entropy,
as can be seen in figure \ref{fig:mtzneut}.
In graph (a) we plot the entropy $S_{c}$ of the cosmological horizon for the neutral MTZ
black holes compared with the entropies of the event horizon ($S{+}$) and cosmological
horizon ($S{++}$) of the Reissner-Nordstr\"om-de Sitter black hole with
the same metric but no scalar field.
The direct proportionality between cosmological horizon entropy and temperature for the
MTZ black holes is readily apparent.
In graph (b) the same entropies are plotted against the mass $M$ of the black hole,
showing how the cosmological horizon entropies decrease as the mass of the black hole
increases, whereas normally entropy increases as black hole mass increases.
The temperature also decreases as mass increases from (\ref{eq:MTZtemp}).
The entropy of the event horizon in Reissner-Nordstr\"om-de Sitter space does increase
as mass increases, but even so the total entropy still decreases as mass increases.

\begin{figure}
\begin{center}
\includegraphics[height=8cm,angle=90]{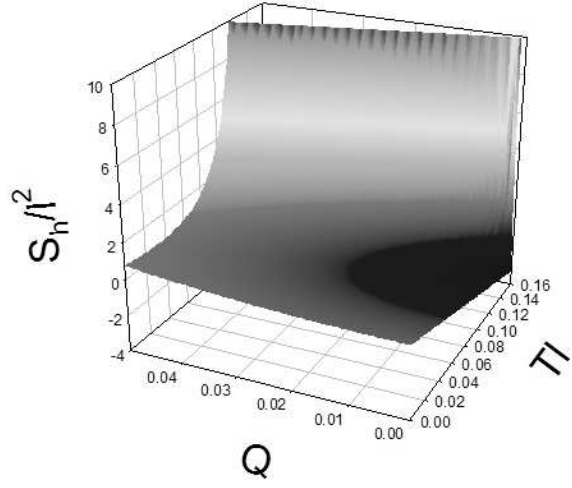}
\caption{Graph of $S_{h}$, the entropy of the event horizon, for
charged MTZ black holes, as a function of temperature $T$ and
charge $Q$.}
\label{fig:mtzSh}
\end{center}
\end{figure}

\begin{figure}
\begin{center}
\includegraphics[width=10cm,angle=270]{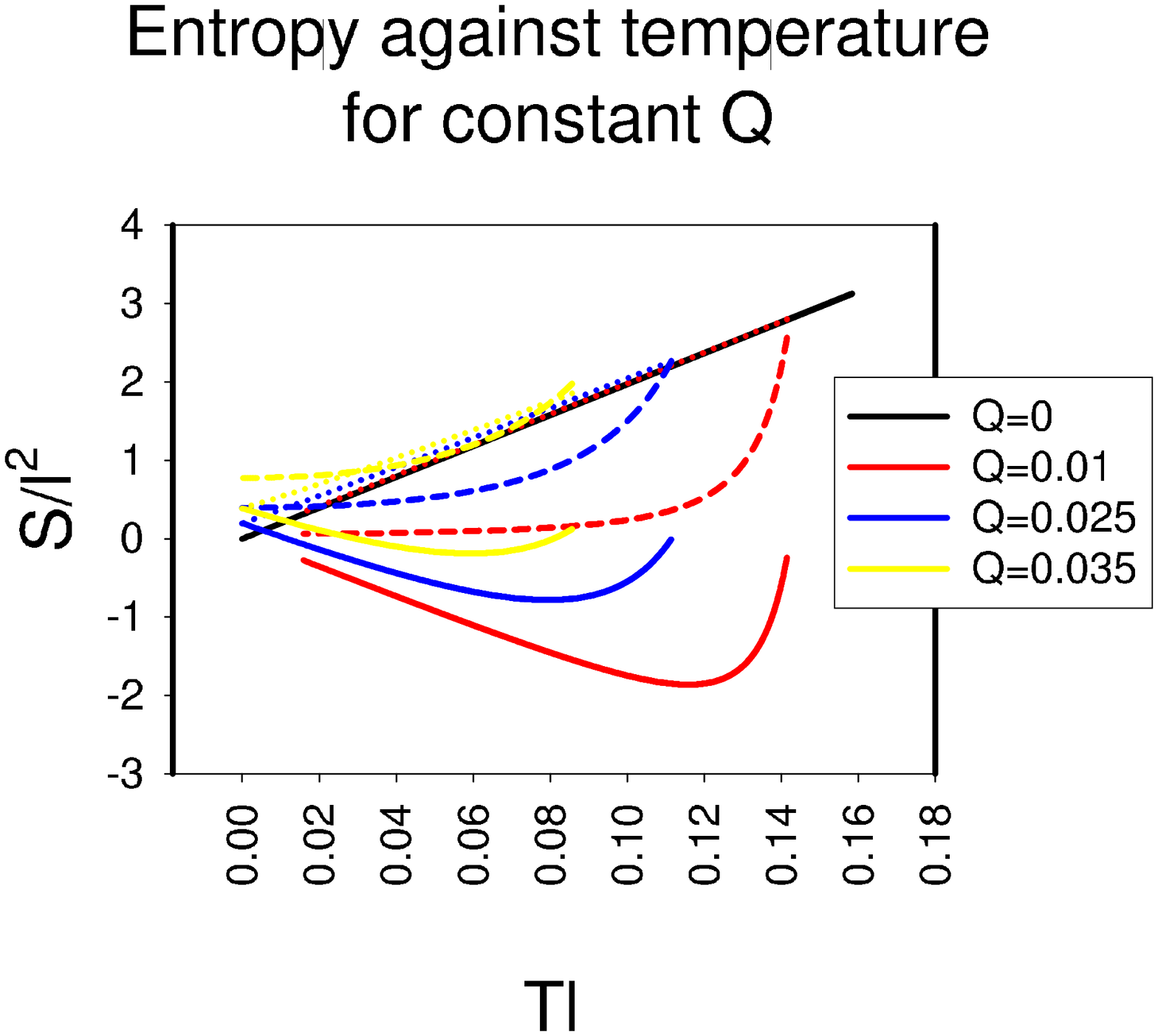}
\caption{Entropy of the charged MTZ black holes as a function of temperature
$T$ for fixed values of the charge $Q$.
The black line corresponds to the neutral MTZ black holes, and represents $S_{c}=-S_{h}$ in
this case.
For the other curves, the solid lines are plots of the event horizon entropy $S_{h}$,
the dotted curves (which all lie very close to the black line) are the cosmological
horizon entropy $S_{c}$, while the dashed curves are the total entropy $S_{h}+S_{c}$.}
\label{fig:mtzcharged}
\end{center}
\end{figure}

Secondly, we turn to the charged MTZ black holes, where the situation is rather more
complicated.
The two horizons still have the same temperature because the metric is unchanged, but
the fact that the scalar field $\phi $ has a different form (\ref{eq:MTZcharged}) alters
the entropy and the event and cosmological horizons no longer have the same entropy.
Instead they are given by
\begin{eqnarray*}
S_{h} & = &
\frac {\pi l^{2}}{2} \left[
1-\frac {2M}{l} - {\sqrt {1-\frac {4M}{l}}}
+ \frac {\Lambda }{36\alpha } \left(
1 - \frac {2M}{l} + {\sqrt {1 - \frac {4M}{l}}}
\right) \right] ;
\nonumber \\
S_{c} & = &
\frac {\pi l^{2}}{2} \left[
1-\frac {2M}{l} + {\sqrt {1-\frac {4M}{l}}}
+ \frac {\Lambda }{36\alpha } \left(
1 - \frac {2M}{l} - {\sqrt {1 - \frac {4M}{l}}}
\right) \right] .
\end{eqnarray*}
Alternatively, it is useful to write the entropies in terms of the mass $M$ and charge
$Q$:
\begin{eqnarray*}
S_{h} & = &
\pi l^{2} \left[
 - {\sqrt {1-\frac {4M}{l}}}
+ \frac {4\pi Q^{2}}{M^{2}} \left(
1 - \frac {2M}{l} + {\sqrt {1 - \frac {4M}{l}}}
\right) \right] ;
\nonumber \\
S_{c} & = &
\pi l^{2} \left[
 {\sqrt {1-\frac {4M}{l}}}
+ \frac {4\pi Q^{2}}{M^{2}} \left(
1 - \frac {2M}{l} - {\sqrt {1 - \frac {4M}{l}}}
\right) \right] ;
\end{eqnarray*}
or, equivalently, as functions of temperature $T$ (\ref{eq:MTZtemp}) and $Q$:
\begin{eqnarray}
S_{h} & = &
2\pi ^{2} \left[
 - l^{3} T
+ 16 Q^{2} \left( 1- 2\pi l T \right) ^{-2}
 \right] ;
\nonumber \\
S_{c} & = &
2\pi ^{2} \left[
l^{3} T
+ 16 Q^{2} \left( 1 - 2\pi l T \right) ^{-2}
\right] .
\label{eq:entropyT}
\end{eqnarray}
In this case the entropy of the event horizon ($S_{h}$) is not always negative, although
it is less than zero for some values of the mass and charge
(see figure \ref{fig:mtzSh}, where we have plotted $S_{h}$ against charge $Q$ and
temperature $T$ (\ref{eq:entropyT})).
On the other hand, the entropy of the cosmological horizon ($S_{c}$) and the total
entropy are always positive:
\begin{displaymath}
S_{h} + S_{c} =
\frac {8\pi ^{2}Q^{2} l^{2}}{M^{2}} \left( 1- \frac {2M}{l} \right) ,
\end{displaymath}
so that the total entropy is zero only when $Q=0$.
We examine whether or not there are phase transitions by plotting,
in figure \ref{fig:mtzcharged}, the entropies as functions of temperature $T$
for fixed values of the charge $Q$ (\ref{eq:entropyT}).
In figure \ref{fig:mtzcharged}, the black line is the neutral black hole cosmological horizon entropy
$S_{c}$ (which is equal to $-S_{h}$ in this case) for comparison purposes.
Each of the solid curves is the event horizon entropy $S_{h}$, the dotted curves (which lie
very close to the black line and so are difficult to distinguish) are the cosmological
horizon entropy $S_{c}$ and the dashed curves are the total entropy $S_{h}+S_{c}$.
From this graph it is clear that $S_{h}$ is mostly negative, but in general becomes positive
for small values of the temperature $T$.
However, both the cosmological horizon entropy and the total entropy can be seen to be
positive.
It can also be seen from figure \ref{fig:mtzcharged} that
the event horizon entropy $S_{h}$ possesses a phase transition, but neither the cosmological
horizon entropy nor the total entropy do.
This can also be checked by calculating the partial derivatives of the entropies
(\ref{eq:entropyT}) with respect to temperature $T$, holding $Q$ and the cosmological
constant $\Lambda $ constant \cite{barlow}.

It is difficult to directly compare the thermodynamics of the charged MTZ black holes with
the usual Reissner-Nordstr\"om-de Sitter black holes.
Although the metrics are the same, the phase space is rather different.
The metric (\ref{eq:MTZcharged}) represents a Reissner-Nordstr\"om-de Sitter black hole
with a charge equal to its mass, however for charged MTZ black holes the charge is
no longer equal to the mass, but is related to it via equation (\ref{eq:QMrel}).
We will examine these issues in more detail elsewhere \cite{barlow}.

\subsection{Thermodynamics of the Asymptotically AdS Black Holes}
\label{sec:adS}

\begin{figure}
\begin{center}
\vspace{1in}
\includegraphics[width=8cm,angle=270]{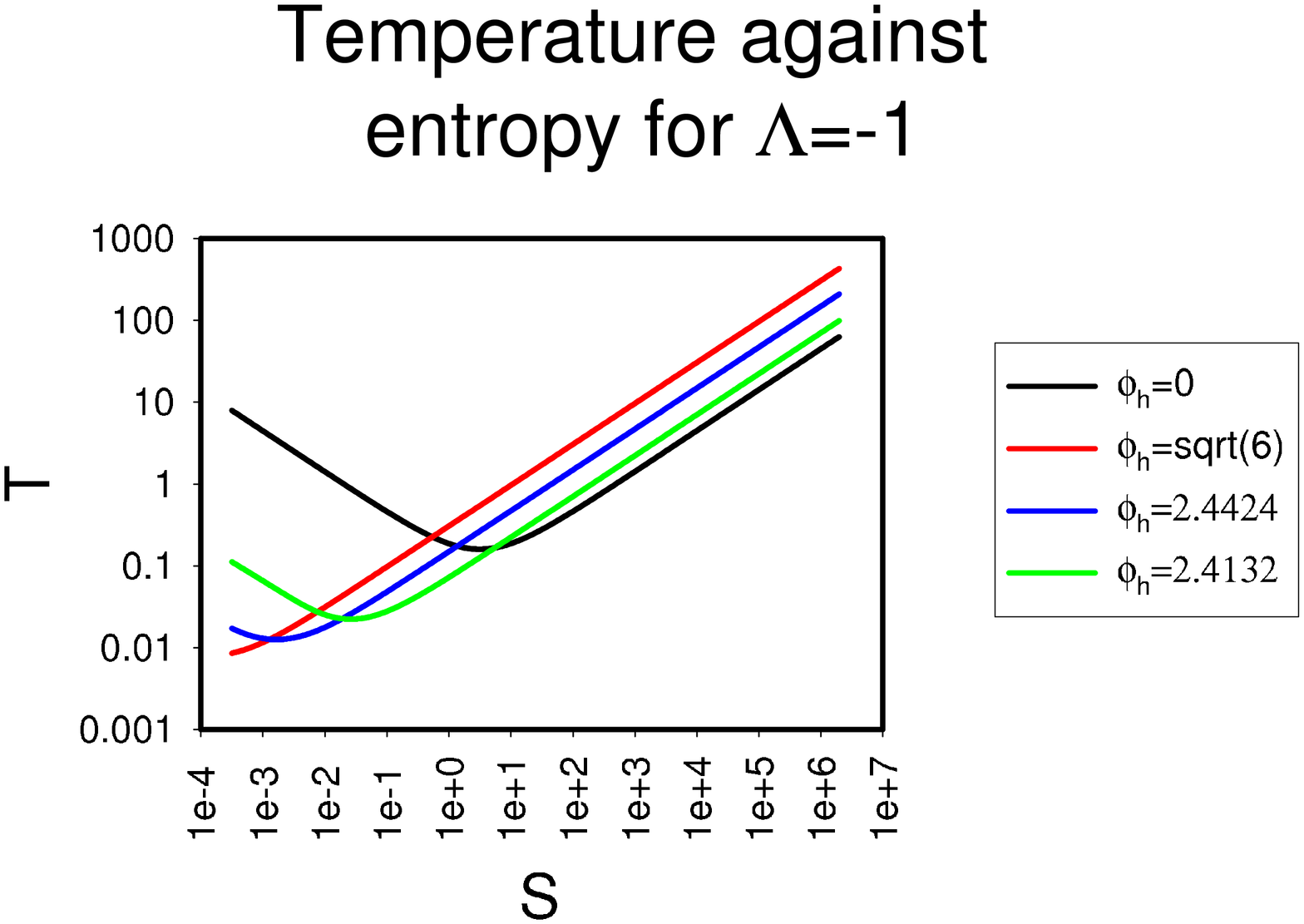}
\caption{Graph of temperature against entropy for the asymptotically anti-de Sitter
black holes with $\Lambda = -1$ and scalar field mass $\mu ^{2}=0$.
Along each curve the value of the scalar field at the event horizon is fixed as shown,
and the radius of the event horizon is varied.}
\label{fig:adstemp}
\end{center}
\end{figure}

In contrast to the MTZ black holes, the thermodynamics of the asymptotically AdS
black holes discussed in section \ref{sec:Lneg} is comparatively simple.
In figure \ref{fig:adstemp} we plot temperature against entropy for solutions with
$\Lambda = -1$ and scalar field mass $\mu ^{2}=0$,
although similar behaviour is observed for other values of $\Lambda $ and $\mu ^{2}$.
Along each curve the value of the conformally coupled scalar field at the event horizon
is fixed as shown, and the radius of the event horizon is varied over several orders of
magnitude.
Note that these solutions are derived by transforming to a minimally coupled system
(\ref{eq:mincscal}) and in order for this transformation to be valid, it must be the
case that $\phi _{h}=\phi (r_{h}) \le {\sqrt {6}} \sim 2.449$.
The black curve in figure \ref{fig:adstemp}, corresponding to $\phi _{h}=0$,
is for ordinary Schwarzschild-anti-de Sitter black holes.
It can be seen that the plots in figure \ref{fig:adstemp}
for other values of $\phi _{h}$ are very similar to those for
Schwarzschild-anti-de Sitter black holes, with a single phase transition, the
Hawking-Page phase transition \cite{hawking}.
Large black holes are thermodynamically stable, while small black holes are thermodynamically
unstable to evaporation.
This similarity to Schwarzschild-anti-de Sitter is not unexpected as holding
$\phi _{h}$ fixed means that the entropy (\ref{eq:entropy}) is proportional to the
usual area law, although the dependence of the temperature on radius with $\phi _{h}$
fixed is not so simple.

\section{Conclusions}
\label{sec:conc}

In this paper we have studied hairy black hole solutions of the Einstein equations
for gravity with a conformally coupled scalar field.
We have reviewed the solutions existing in the literature when the cosmological constant
is zero, positive or negative and discussed the stability of these solutions
under linear, spherically symmetric perturbations, which can be summarized in the following
table \cite{harper}:
\newline
\begin{center}
\begin{tabular}{cc}
\hline
$\Lambda =0$  & Unstable hair
\\
$\Lambda >0$  & Unstable hair
\\
$\Lambda <0$  & Stable hair
\\
\hline
\end{tabular}
\end{center}
We have also studied how the presence of the non-minimal coupling of the scalar field to the
space-time curvature affects the classical thermodynamics of these black holes, in
particular their entropy (\ref{eq:entropy}).
Some surprising features emerge:
\begin{enumerate}
\item
When the cosmological constant is zero, the BBMB solution has zero temperature but
an entropy which is formally infinite.
Attempts in the literature to understand this have led to the conclusion that
conventional thermodynamics cannot be applied to this black hole \cite{zaslavskii}.
\item
When the cosmological constant is positive, the thermodynamics of the MTZ solutions is rather
complex \cite{barlow}.
The event horizon usually has negative entropy, but for the neutral black holes this is
completely cancelled by the entropy of the cosmological horizon, leading to a zero total
entropy.
For the charged black holes, this cancellation does not occur and the total entropy is
positive.
\item
For negative cosmological constant, the entropy-temperature relation is qualitatively
similar to that for ordinary Schwarzschild-anti-de Sitter black holes, despite the
conformal coupling of the scalar field.
\end{enumerate}

\section*{Acknowledgements}
I would like to thank the organizers of the workshop on ``Dynamics and Thermodynamics
of Black Holes and Naked Singularities'' for a most enjoyable and stimulating meeting.
Much of the work in this talk was done jointly with the following collaborators:
Anne-Marie Barlow, Dan Doherty, Tom Harper, Paul Thomas and Phil Young.
This work was supported by a grant from the Nuffield Foundation, grant reference number
NUF-NAL/02, and the work of PY is supported by a studentship from the EPSRC.


\begin{thebibliography}{00}

\bibitem{bek74}
J.~D. Bekenstein,
{\em Ann. Phys.} {\bf 82}, 535--547 (1974).

\bibitem{bek75}
J.~D. Bekenstein,
{\em Ann. Phys.} {\bf 91}, 75--82 (1975).

\bibitem{bbm}
N.~M. Bocharova, K.~A. Bronnikov, and V.~N. Mel'nikov,
{\em Vestnik Moskov. Univ. Fizika}
{\bf 25}, 706--709 (1970).

\bibitem{heuslerbook}
M. Heusler, {\em Black Hole Uniqueness Theorems}
(Cambridge University Press, Cambridge, 1996).

\bibitem{saa1}
A. Saa,
{\em J. Math. Phys.} {\bf 37}, 2346--2351 (1996).

\bibitem{saa2}
A. Saa,
{\em Phys. Rev.} {\bf D53}, 7377--7380 (1996).

\bibitem{martinez}
C. Martinez, R. Troncoso, and J. Zanelli,
{\em Phys. Rev.} {\bf D67}, 024008 (2003).

\bibitem{ewconf}
E. Winstanley,
{\em Found. Phys.} {\bf 33}, 111--143 (2003).

\bibitem{ashtekar}
A. Ashtekar, A. Corichi and D. Sudarsky,
{\em Class. Quantum Gravity} {\bf 20}, 3413--3425 (2003).

\bibitem{maeda}
K-I. Maeda,
{\em Phys. Rev.} {\bf D39}, 3159--3162 (1989).

\bibitem{zaslavskii}
O.~B. Zaslavskii,
{\em Class. Quantum Gravity} {\bf 19}, 3783--3797 (2002).

\bibitem{barlow}
A.-M. Barlow, D. Doherty and E. Winstanley,
work in preparation.

\bibitem{bek96}
J.~D. Bekenstein,
 in: I.~M. Dremin and A.~M. Semikhatov (eds.): {\em Proceedings of the
  Second International Sakharov Conference on Physics, Moscow, Russia, 20-23
  May 1996}. pp. 216--219 (World Scientific, Singapore, 1997).

\bibitem{zannias}
D. Sudarsky and T. Zannias,
{\em Phys. Rev.} {\bf D58}, 087502 (1998).

\bibitem{bronnikov}
K.~A. Bronnikov and Y.~N. Kireyev,
{\em Phys. Lett.} {\bf 67A}, 95--96 (1978).

\bibitem{xanth}
B.~C. Xanthopoulos and T. Zannias,
{\em J. Math. Phys.} {\bf 32}, 1875--1880 (1991).

\bibitem{mayo}
A.~E. Mayo and J.~D. Bekenstein,
{\em Phys. Rev.} {\bf D54}, 5059--5069 (1996).

\bibitem{bek72}
J.~D. Bekenstein,
{\em Phys. Rev.} {\bf D5}, 1239--1246 (1972).

\bibitem{bek95}
J.~D. Bekenstein,
{\em Phys. Rev.} {\bf D51}, R6608--R6611 (1995).

\bibitem{heusler95}
M. Heusler,
{\em Class. Quantum Gravity} {\bf 12}, 779--789 (1995).

\bibitem{heusler92}
M. Heusler and N. Straumann,
{\em Class. Quantum Gravity} {\bf 9}, 2177--2189 (1992).

\bibitem{sudarsky}
D. Sudarsky,
{\em Class. Quantum Gravity} {\bf 12}, 579--584 (1995).

\bibitem{pena}
I. Pe\~na and D. Sudarsky,
{\em Class. Quantum Gravity} {\bf 18}, 1461--1474 (2001).

\bibitem{harper}
T.~J.~T. Harper, P.~A. Thomas, E. Winstanley and P.~M. Young,
``Instability of a four-dimensional de Sitter black hole with a conformally
coupled scalar field'', Preprint {\em gr-qc/0312104} (2003),
to appear in {\em Phys. Rev. D}.

\bibitem{toriidS}
T. Torii, K. Maeda, and M. Narita,
{\em Phys. Rev.} {\bf D59}, 064027 (1999).

\bibitem{iyer}
V. Iyer and R.~M. Wald,
{\em Phys. Rev.} {\bf D50}, 846--864 (1994).

\bibitem{fatibene}
L. Fatibene, M. Ferraris, M. Francaviglia and M. Raiteri,
{\em Ann. Phys.} {\bf 275}, 27--53 (1999).

\bibitem{mellor1}
F. Mellor and I.~G. Moss,
{\em Class. Quantum Gravity} {\bf 6}, 1379--1385 (1989).

\bibitem{mellor2}
F. Mellor and I.~G. Moss,
{\em Phys. Lett.} {\bf B222}, 361--363 (1989).

\bibitem{hawking}
S.~W. Hawking and D.~N. Page,
{\em Commun. Math. Phys.} {\bf 87}, 577--588  (1983).


\end{thebibliography}
\end{document}